\documentclass{aa}
\usepackage{txfonts}
\usepackage{natbib}
\bibpunct{(}{)}{;}{a}{}{,}    
\usepackage{graphicx}
\usepackage{subfigure}
\usepackage[breaklinks=true, colorlinks=true, urlcolor=magenta, citecolor=blue, linkcolor=blue]{hyperref}

\begin{document}

\title{Identifying reliable periods in 2MASS J09213414-5939068, IGR J16167-4957, and V667 Pup}

\author{
        Arti Joshi \inst{\ref{PUC} 
   \thanks{E-mail: ajoshi@astro.puc.cl, aartijoshiphysics@gmail.com} }
            }
\institute{
Institute of Astrophysics, Pontificia Universidad Católica de Chile, Av. Vicuña MacKenna 4860, 7820436, Santiago, Chile \label{PUC}
}

\abstract{%
Detailed timing analyses of three cataclysmic variables, namely 2MASS J09213414$-$5939068, IGR J16167$-$4957, and V667 Pup are carried out using the long-baseline and high-cadence optical photometric data from the Transiting Exoplanet Survey Satellite (\textit {TESS}). Periods of 908.12$\pm$0.05 s and 990.10$\pm$0.06 s are observed in the optical variation of 2MASS J09213414$-$5939068 that were not found in earlier studies and appear to be probable spin and beat periods of the system, respectively. The presence of multiple periods at spin, beat, and other sidebands indicates that 2MASS J09213414$-$5939068 likely belongs to an intermediate polar class of magnetic cataclysmic variables that seems to be accreted via a disc-overflow mechanism. Clear evidence of a period of 582.45$\pm$0.04 s is found during the \textit {TESS} observations of IGR J16167$-$4957, which can be interpreted as the spin period of the system. Strong modulation at this frequency supports its classification as an intermediate polar, where accretion may primarily be governed by a disc. The dominance of the spin pulse unveils the disc-fed dominance accretion in V667 Pup, but the detection of the previously unknown beat period of 525.77$\pm$0.03 s suggests that a portion of the material is also accreted through a stream. Moreover, the double-peaked structure observed in the optical spin pulse profile of V667 Pup suggests the possibility of a two-pole accretion geometry, where each pole accretes at a different rate and is separated by 180$^\circ$.}


\titlerunning{Identifying reliable periods in three cataclysmic variables}
\authorrunning{Joshi et~al.}
\maketitle


\section{Introduction}
\label{sec:intro}
Cataclysmic variables (CVs) are semi-detached binary systems in which a primary white dwarf (WD) accretes material from a Roche-lobe-filling late-type secondary star. When the WD magnetic field is $<$ 10$^{6}$ G, the material transferred from the secondary flows through the inner Lagrangian point and orbits the primary, forming an accretion disc \citep{Warner95}. However, when the WD is highly magnetised ($\sim$ 10$^{6-9}$ G), the accretion disc is either fully \citep[][e.g. known as AM Herculis; a polar]{Cropper90} or partially disrupted \citep[][e.g., known as DQ Herculis; an intermediate polar (IP)]{Patterson94}.

The  magnetic field strength of the WD plays a vital role in controlling the motion of the accretion flow within the effective magnetospheric radius. In IPs, the magnetic field strength of the WD (typically less than 10$^7$ G) causes the accreting material to form an accretion disc. However, at a certain point, the magnetic pressure exceeds the ram pressure, leading to the disruption of the disc. The material then flows along the magnetic field lines. The matter in the accretion column impacts the poles of WDs with supersonic velocities, which leads to the formation of strong shocks that emit cyclotron (optical/infrared) and thermal bremsstrahlung (X-ray) radiation \citep{Aizu73}. The majority of the IPs peaks at the hard luminosity of 10$^{33}$$-$$10^{34}$ erg s$^{-1}$, but there is also evidence of low-luminosity IPs. These are mostly shorter-period systems below the period gap of 2-3 h, and they have a luminosity of $\sim$ 10$^{30}$$-$$10^{32}$ erg s$^{-1}$ \citep{Schwope18, deMartino20}. 

\begin{table*}
\normalsize
\caption{Log of the \textit {TESS} observations with start and end times in calendar date.\label{tab:obslog}}
\setlength{\tabcolsep}{0.05in}
\centering
\begin{tabular}{@{}ccccccccc@{}}
\hline\\
Object  & Sector  & Start Time & End Time & Total observing days     \\
\hline        \\                                      2MASS J09213414$-$5939068    &  09      &2019-02-28T18:08:25.5 & 2019-03-25T23:26:48.1 & 25.2 \\
                           &  10      &2019-03-26T22:30:47.7 & 2019-04-22T04:22:31.0 & 26.2 \\
IGR J16167$-$4957            &  12      &2019-05-21T11:52:43.3 & 2019-06-18T09:30:37.7 & 27.9 \\
V667 Pup                   &  34      &2021-01-14T06:29:54.5 & 2021-02-08T13:39:26.4 & 25.2\\    
\hline
\end{tabular}
\end{table*}
Intermediate polars are asynchronous systems, where the spin period ($P_\omega$) of the WD is typically shorter than the orbital period ($P_\Omega$) of the binary system. The optical and X-ray signals in IPs are modulated on the spin period, orbital period, beat period ($P_\omega$$_{-}$$_\Omega$), and other sidebands. The multiple periodicities in the X-ray and optical bands is a distinguishing feature of these systems and are caused by intricate interactions between the spin and orbital modulations, either as a result of changes in the accretion rate or as a result of the reprocessing of primary radiation by other components of the system. Thus, an essential diagnostic tool for establishing the IP nature of a possible member of this class are multiple periodic components. Moreover, three different accretion mechanisms are thought to occur in IPs: disc-fed, stream-fed (disc-less), and disc-overflow. These mechanisms depend upon the magnetic field strength of the WD, on the mass accretion rate, and on the binary orbital separation. In these scenarios, the mode of accretion can be identified through the orbital, spin, beat, and sideband frequencies. In disc-fed accretion, the inner edge of the accretion disc is truncated at the magnetospheric radius, which results in the formation of so-called accretion curtains near the magnetic poles of the WD \citep{Rosen88}. A strong modulation occurs at the spin frequency of the WD in disc-fed accretion \citep{Kim95, Norton96}. In the disc-less or stream-fed accretion, the high magnetic field of the WD prevents the formation of a disc, and infalling material flows along the magnetic field lines to the pole caps \citep{Hameury86}. Modulation at the beat frequency \citep{Hellier91, Wynn92} is a true indicator of disc-less accretion. According to \cite{Wynn92}, an asymmetry between the magnetic poles can also lead to a modulation at the spin frequency in addition to the beat frequency when accretion occurs via the accretion stream. Then the 2$\omega-\Omega$ frequency in X-rays significantly contributes to confirming the accretion mode as stream-fed, along with the sometimes dominant $\Omega$ component, $\omega-\Omega$, and $\omega$. In the disc-overflow accretion \citep{Lubow89, Armitage96}, a combination of both disc-fed and stream-fed accretions can occur, but a part of the accretion stream skips the disc and directly interacts with the WD magnetosphere \citep{Hellier89, King91}. Modulations at both spin and beat frequencies are expected to occur, and the main difference lies in the varying power and amplitude between the two \citep{Hellier91, Hellier93}. 

In this paper, I present detailed optical analyses of three CVs, namely, 2MASS J09213414$-$5939068, IGR J16167$-$4957, and V667 Pup. The light-curve morphologies, spin period, other sideband periods, and the specific classification of 2MASS J09213414$-$5939068 are still unknown. They are essential for investigating the nature of this system, however. Moreover, the absence of a spin signal in IGR J16167$-$4957 and the lack of beat modulation in V667 Pup led to revisit their properties, which are vital for revealing the exact nature and mode of accretion in these systems. I therefore  present precise and uninterrupted long \textit {TESS} optical observations to unveil the nature of these systems. A summary of the available information for these sources is given below.

2MASS J09213414$-$5939068 (hereafter J0921) was selected as a CV based on the detection of H$\alpha$ emission \citep{Pretorius08}. \cite{Pretorius08} spectroscopic studies revealed a strong He II ($\lambda$4686 \AA) emission line, as well as the Balmer and He I emission lines, overlaid on a blue continuum. However, photometric observations exhibited no modulations other than flickering. Based on time-resolved spectroscopic data, a period of 3.041$\pm$0.009 hr was determined and assumed to be the orbital period of the system.  
 
IGR J16167$-$4957 (hereafter J1616) was discovered as a hard X-ray source by \cite{Barlow06}, and \citep{Tomsick06} concluded it was not one of the high-mass X-ray binaries. It was identified as a CV by \cite{Masetti06} based on the optical spectral features. Later, \cite{Pretorius09} determined its orbital period as 5.004$\pm$0.005 hr using the time-resolved spectroscopic observations. The orbital period was not detected in their high-speed photometric observations. A signal of approximately 585 s was observed for half of the observing time for the one-epoch light curve, but it was absent in their other observational epochs, leading them to hypothesise that it might be a quasi-periodic signal. Subsequently, the 
 Rossi X-ray Timing Explorer (\textit {RXTE}) spectrum  led to the suggestion that J1616 is likely to be classified as an IP \citep{Butters11}. The lack of an X-ray spin period and a period of 585 s means that its status as an unclassified system remains, however.

V667 Pup has frequently been mentioned in the Astronomer's Telegrams over the course of several months early in 2006. It was first identified as a CV by \cite{Ajello06} based on Neil Gehrels Swift Observatory (\textit {Swift}) observations. \cite{Masetti06} identified two nearby objects, the brighter of which exhibited a spectrum typical of a normal G/K type Galactic star. However, the fainter object was identified as the true optical counterpart with the detection of spectral features, including emission lines of $H\alpha$, $H\beta$, HeI ($\lambda$5875 \AA), and HeII ($\lambda$4686 \AA). These signatures were concluded to be those of an accretion disc in a low-mass X-ray binary. Later, \cite{Patterson06} reported a stable spin pulsation at a period of $\sim$ 512.4 s using optical photometry and classified it as an IP. The optical counterpart of V667 Pup was further observed by \cite{Marsh06}, who detected spin pulsations identical to those found by \cite{Patterson06}. In the same year, \cite{Torres06} reinforced its classification as a probable IP with the detection of Balmer emission lines from H$\alpha$ to H$\gamma$, including HeII ($\lambda$4686 \AA), HeI ($\lambda$5876 \AA), HeI ($\lambda$6678 \AA), and the Bowen blend in emission. Upon revisiting the \textit {Swift} data, \cite{Wheatley06} detected an X-ray pulsation consistent with the proposed spin period of $\sim$ 512 s. Based on radial velocity measurements of the H$\alpha$ emission lines, \cite{Thorstensen06} derived an orbital period of 5.604$\pm$0.019 hr, which was subsequently refined by \cite{Thorstensen13} as 5.611$\pm$0.005 hr. \cite{Butters07} reported an X-ray spin period of 512.4$\pm$0.3 s based on the \textit {RXTE} observations, which supported the classification of this object as an IP. Later, precise photometric measurements of this object were reported by \cite{Thorstensen13}, who revealed fluctuations of about 0.05 mag for V667 Pup, while the companion exhibited somewhat smaller variations. Moreover, their observed spectrum revealed relatively weak emission lines; HeII ($\lambda$4686 \AA) displayed a strength that was comparable to that of the H$\beta$ emission line. At a period of 5.61 hr, their analysis revealed a prominent modulation in the H$\alpha$ emission line velocities.

The paper is organised as follows:  The observations and data reduction are summarised in the next section. Section \ref{sec:analysis} contains analyses and the results. Finally, the discussion and conclusions are presented in Sections \ref{sec:diss} and \ref{sec:conc}, respectively.


\begin{figure*}
\centering
\subfigure[]{\includegraphics[width=0.82\textwidth]{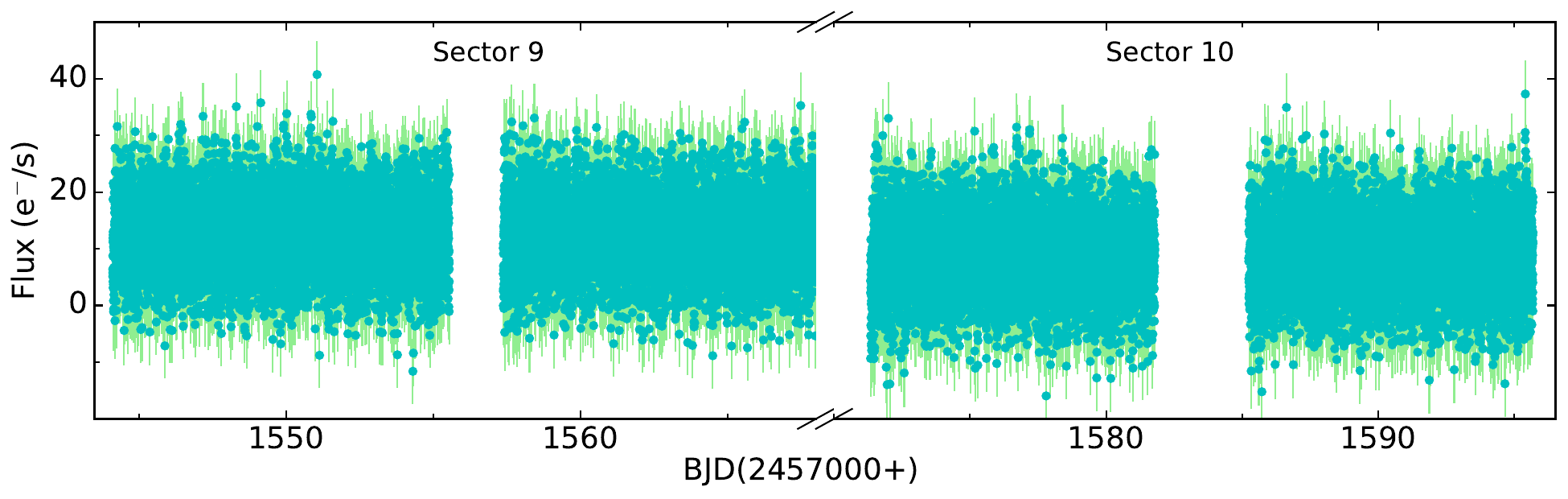}\label{fig:tesslc_J0921}}
\subfigure[]{\includegraphics[width=0.82\textwidth]{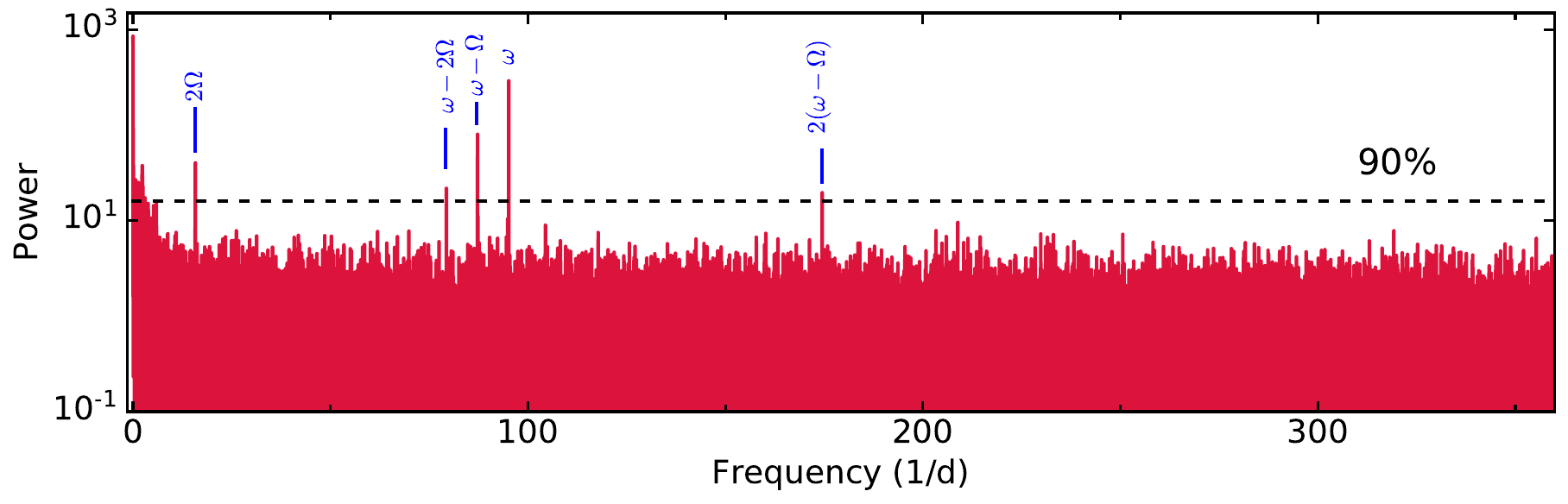}\label{fig:tessps_J0921}}
\caption{(a) \textit{TESS} light curves of J0921 to reveal the brightness variations in sectors 9 and 10, while (b) LS power spectrum obtained from the combined \textit{TESS} observations of sectors 9 and 10. The significant frequencies observed in the power spectrum lie above the 90\% confidence level (represented by the dashed horizontal black line) and are marked.} 
\end{figure*}

\begin{figure*}
\centering
\subfigure[]{\includegraphics[width=0.85\textwidth]{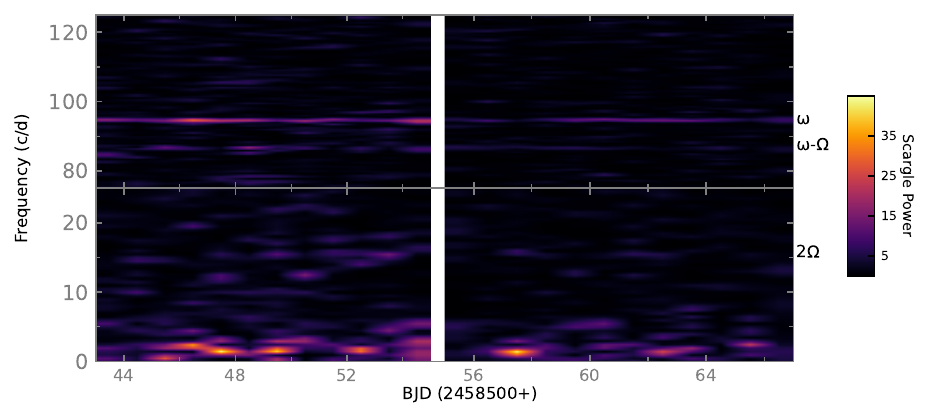}\label{fig:1daypssec9_J0921}}
\subfigure[]{\includegraphics[width=0.85\textwidth]{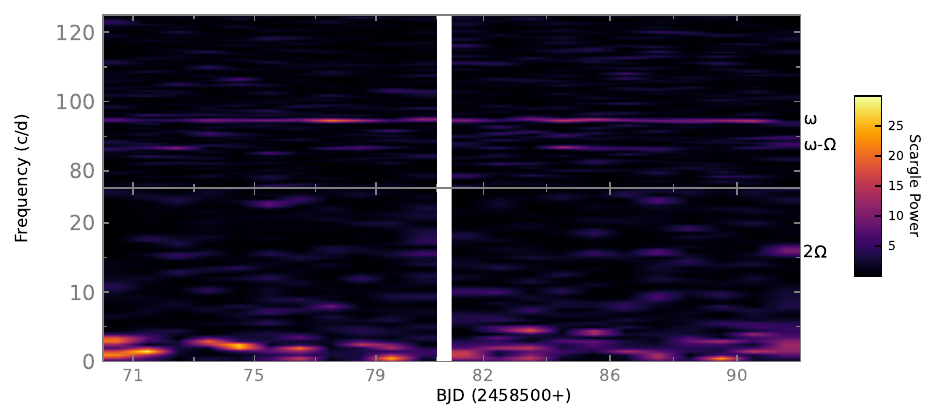}\label{fig:1daypssec10_J0921}}
\caption{One-day time-resolved \textit {TESS} power spectra of J0921 near the $2\Omega$ and $\omega$ frequency region for (a) sector 9 and (b) sector 10.} 
\end{figure*}

\begin{figure*}
\centering
\includegraphics[width=0.7\textwidth]{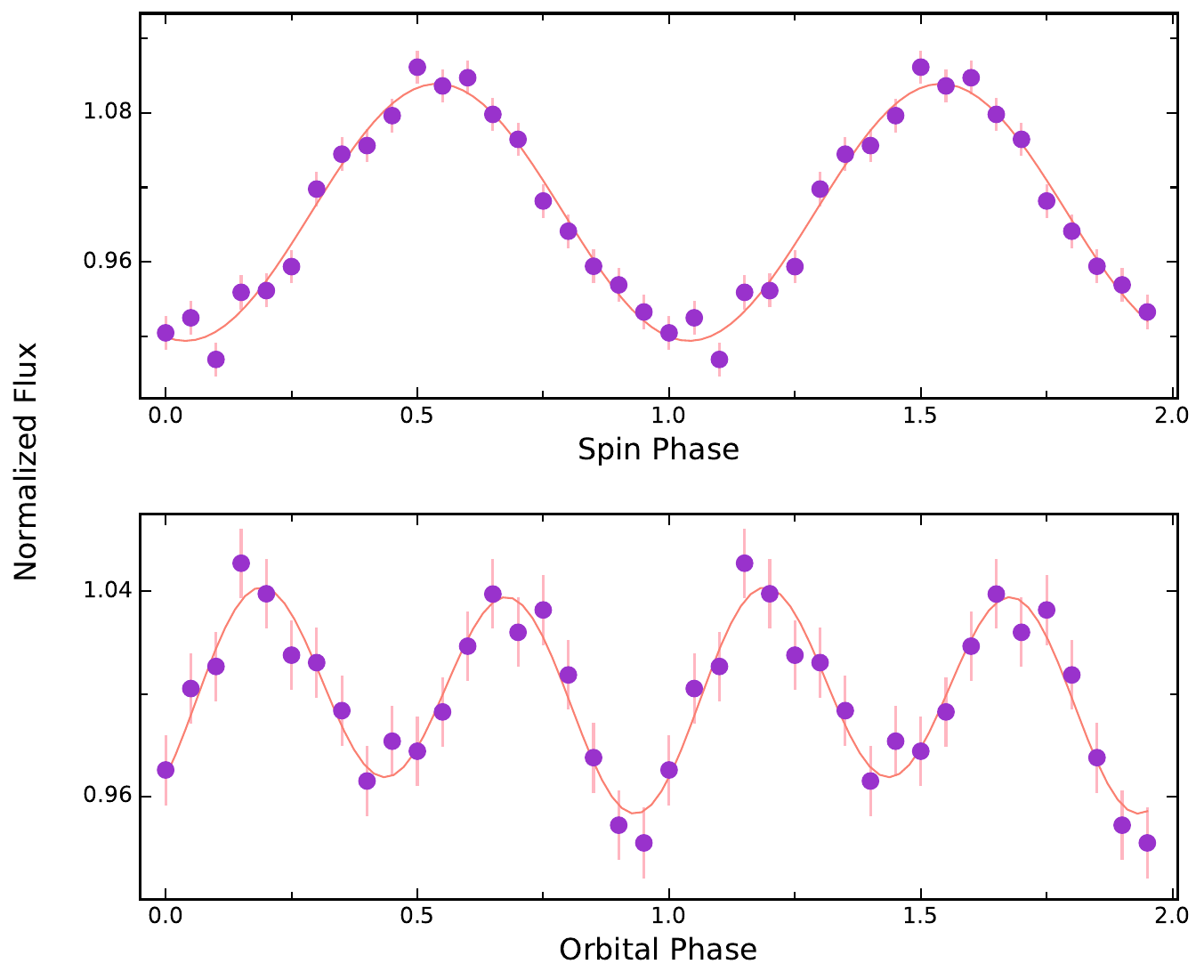}
\caption{Spin (top) and orbital (bottom) phase-folded \textit {TESS} light curves of J0921 obtained from the combined data of sectors 9 and 10.} 
\label{fig:tessorbspinflc_J0921}
\end{figure*}

\begin{figure*}
\centering
\subfigure[]{\includegraphics[width=0.48\textwidth]{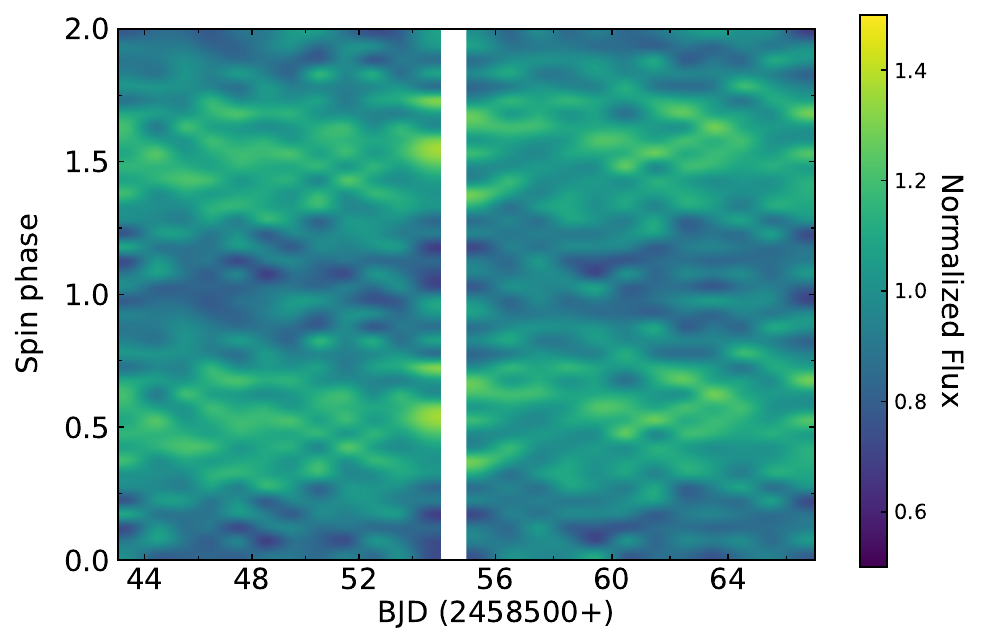}\label{fig:1dayspinflcsec9_J0921}}
\subfigure[]{\includegraphics[width=0.48\textwidth]{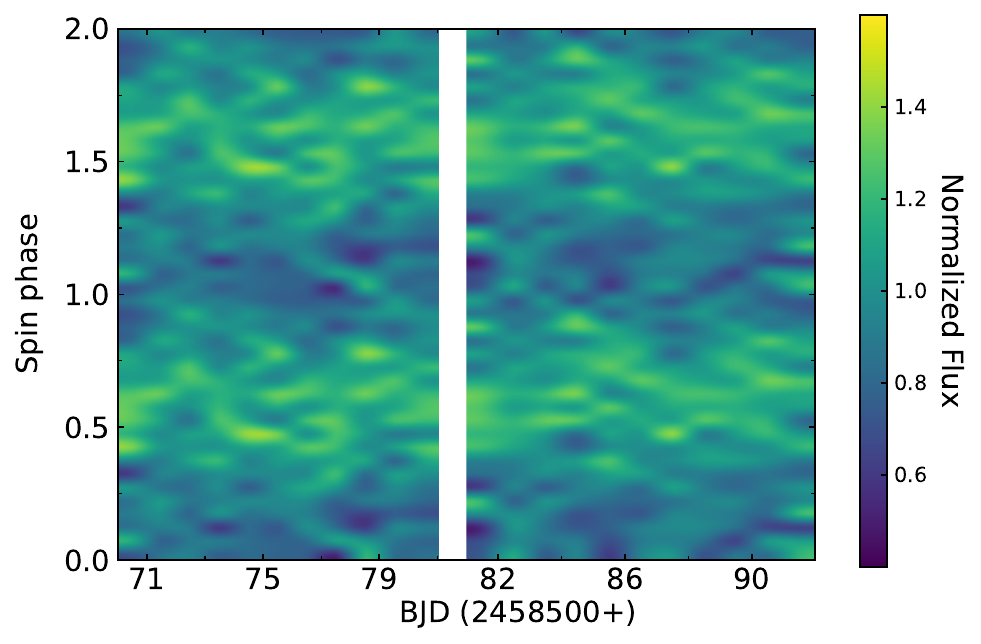}\label{fig:1dayspinflcsec10_J0921}}
\caption{Evolution of the spin-pulse profiles during one-day \textit{TESS} observations of J0921 for (a) sector 9 and (b) sector 10.} 
\end{figure*}

\section{Observations and data reduction}
\label{sec:obs}
The archival data used in this study were obtained from observations by \textit {TESS}  for all three systems. \textit {TESS} has four small telescopes with four cameras with a field-of-view of 24$\times$24 degree$^2$ each, which are aligned to cover 24 $\times$ 96 degree strips of the sky called `sectors' \citep[see][for details]{Ricker15}. The \textit {TESS} bandpass covers the wavelength range from 600 nm to 1000 nm, with an effective wavelength of 800 nm. A detailed log for the \textit {TESS}  observations of all three sources is given in Table \ref{tab:obslog}. The data are available at the Mikulski Archive for Space Telescopes (MAST) data archive\footnote{\textcolor{magenta}{\url{https://mast.stsci.edu/portal/Mashup/Clients/Mast/Portal.html}}} with the identification numbers TIC 387201824, TIC 411381210, and TIC 750058684 for J0921, J1616, and V667 Pup, respectively. The cadence for each source was 2 min. The \textit {TESS} pipeline provides two flux values: simple aperture photometry (SAP), and pre-search data-conditioned SAP (PDCSAP). The PDCSAP light curve attempts to remove instrumental systematic variations by fitting and removing the signals that are common to all stars on the same CCD\footnote{\textcolor{black}{see section 2.1 of the \textit{TESS} archive manual available at \textcolor{magenta}{\url {https://outerspace.stsci.edu/display/TESS/2.0+-+Data+Product+Overview}}}}. By employing this approach, aperiodic variability might be removed from the data. PDCSAP also corrects for the amount of flux captured by the photometric aperture and for crowding from known nearby stars, while SAP does not. To confirm that the PDCSAP data retain  all aperiodic variability, a periodogram analysis was performed on all sources, considering both SAP and PDCSAP data.
The PDCSAP and SAP light curves both showed the same short-period variability, indicating that the PDCSAP data are reliable and can be used for the further analysis. Anomalous event-related data displayed quality flags above 0 in the FITS file structure, and I therefore selected only the data with a QUALITY flag = 0.

\begin{table*}
\small
\centering
\caption{Periods corresponding to dominant peaks in the LS power spectra of J0921, J1616, and V667 Pup.}\label{tab:ps}
\setlength{\tabcolsep}{0.05in}
\begin{tabular}{lccccccccc}
\hline\\
Object              && Sector & \multicolumn{6}{c}{Periods} \\
\cline{4-10}\\
                &&                   &   $P_{\Omega}$  &  $P_{2\Omega}$ &  $P_{(\omega - 2\Omega)}$  & $P_{(\omega - \Omega)}$ &$P_{\omega}$ & $P_{2\omega}$ &   $P_{2{(\omega - \Omega)}}$  \\
	        &&                   &  (hr)          &(hr)           & ( s )                               & (s)                               & ( s )       & (s)           & (s)       \\
\hline\\
J0921           && 9                 &$..$            &$1.518\pm0.001$&$1088.80\pm0.14$ &$990.21\pm0.12$  & $907.98\pm0.11$ & $..$            & $495.13\pm0.03$ \\
                && 10                &$..$            &$1.518\pm0.001$&$1088.79\pm0.14$ &$990.25\pm0.12$  & $908.01\pm0.11$ & $..$            & $495.11\pm0.03$ \\
                && Combined$^\dagger$&$..$            &$1.519\pm0.001$&$1088.96\pm0.07$ &$990.10\pm0.06$  & $908.12\pm0.05$ & $..$            & $495.10\pm0.01$ \\
J1616           && 12                &$4.99\pm0.01$   &$..$           &$..$             &$..$             & $582.45\pm0.04$ & $..$            &$..$ \\
V667 Pup        && 34                &$5.58\pm0.01$   &$2.79\pm0.01$  &$..$             &$525.77\pm0.03$  & $512.40\pm0.03$ & $256.19\pm0.01$ & $..$ \\
\hline
\end{tabular}
\\~\\
{$\dagger$ represents the periods derived from the combined \textit {TESS} data of sectors 9 and 10.}
\end{table*}

\begin{figure*}
\centering
\subfigure[]
{\includegraphics[width=0.66\textwidth]{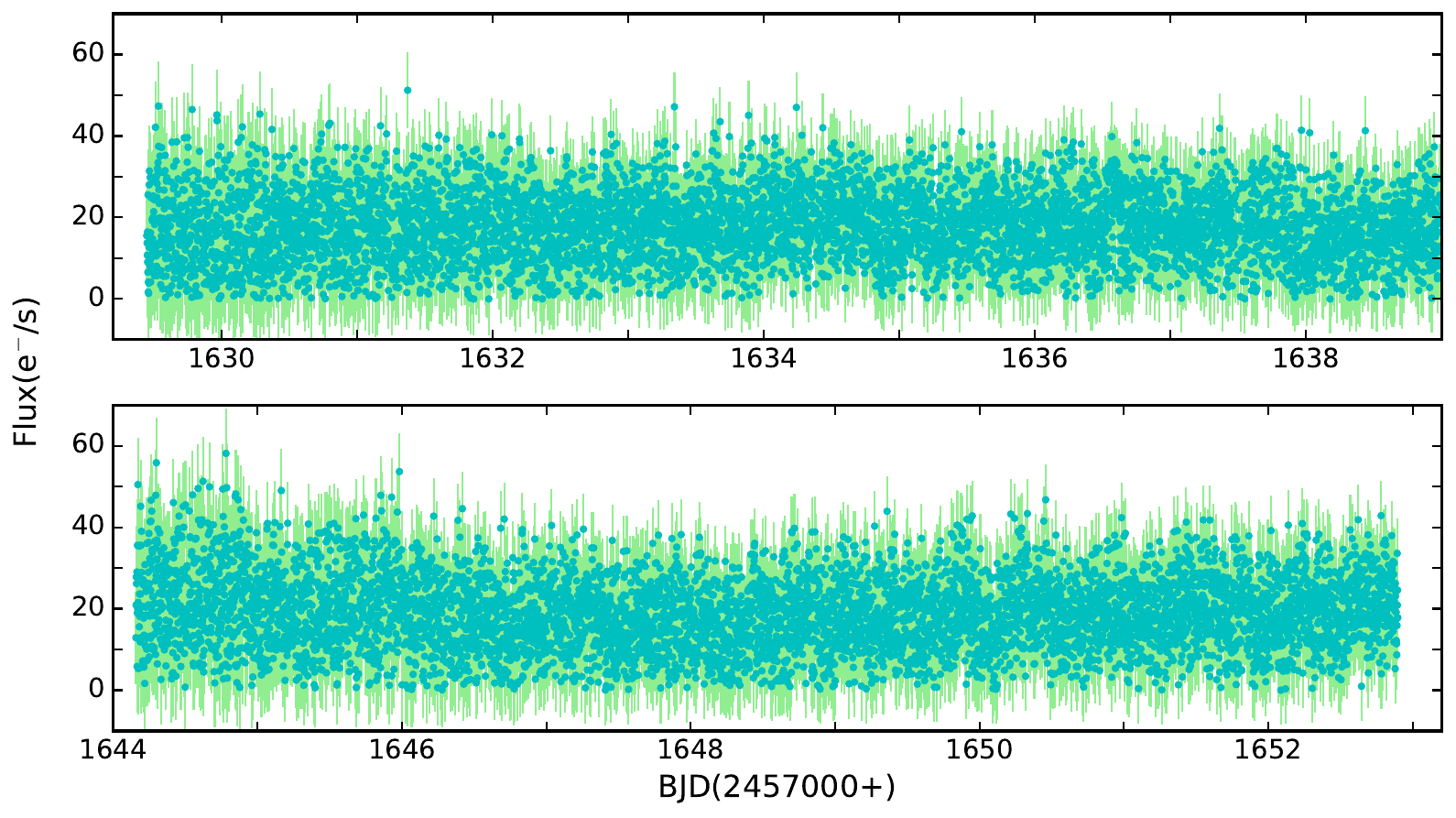}\label{fig:tesslc_J1616}}
\subfigure[]{\includegraphics[width=0.68\textwidth]{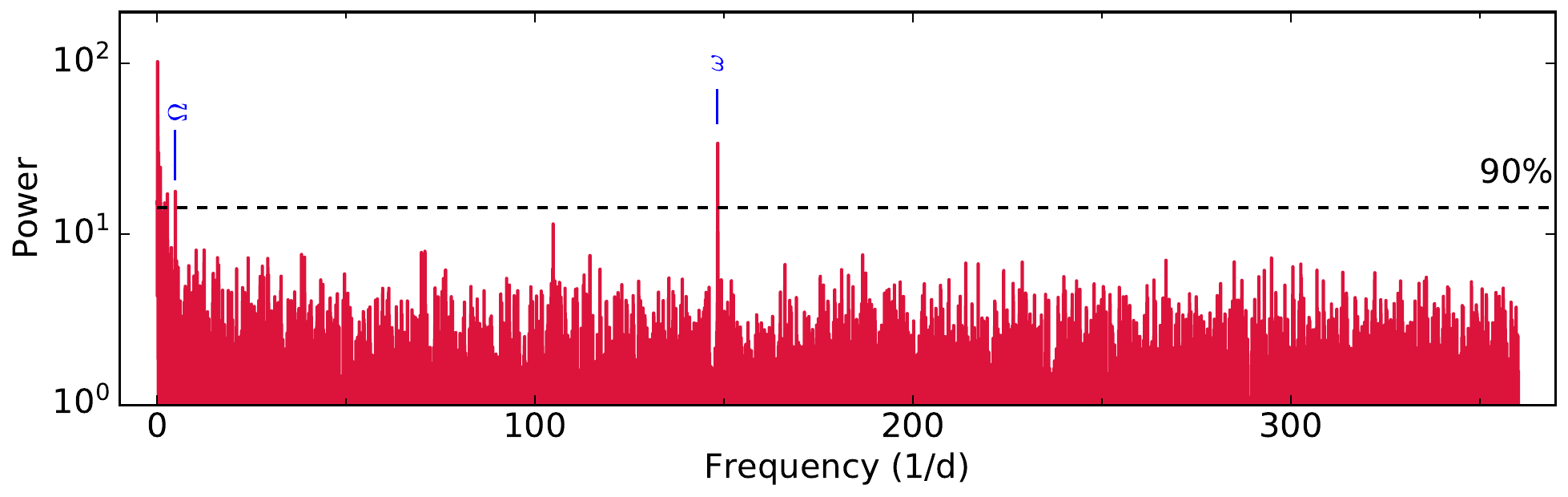}\label{fig:tessps_J1616}}
\caption{(a) \textit{TESS} light curve  
 to reveal the brightness variations in J1616, and (b) LS power spectrum obtained from the \textit {TESS} observations of J1616. The significant frequencies observed in the power spectrum lie above the 90\% confidence level (represented by the dashed horizontal black line) and are marked.} 
\end{figure*}

\begin{figure*}
\centering
\includegraphics[width=0.67\textwidth]{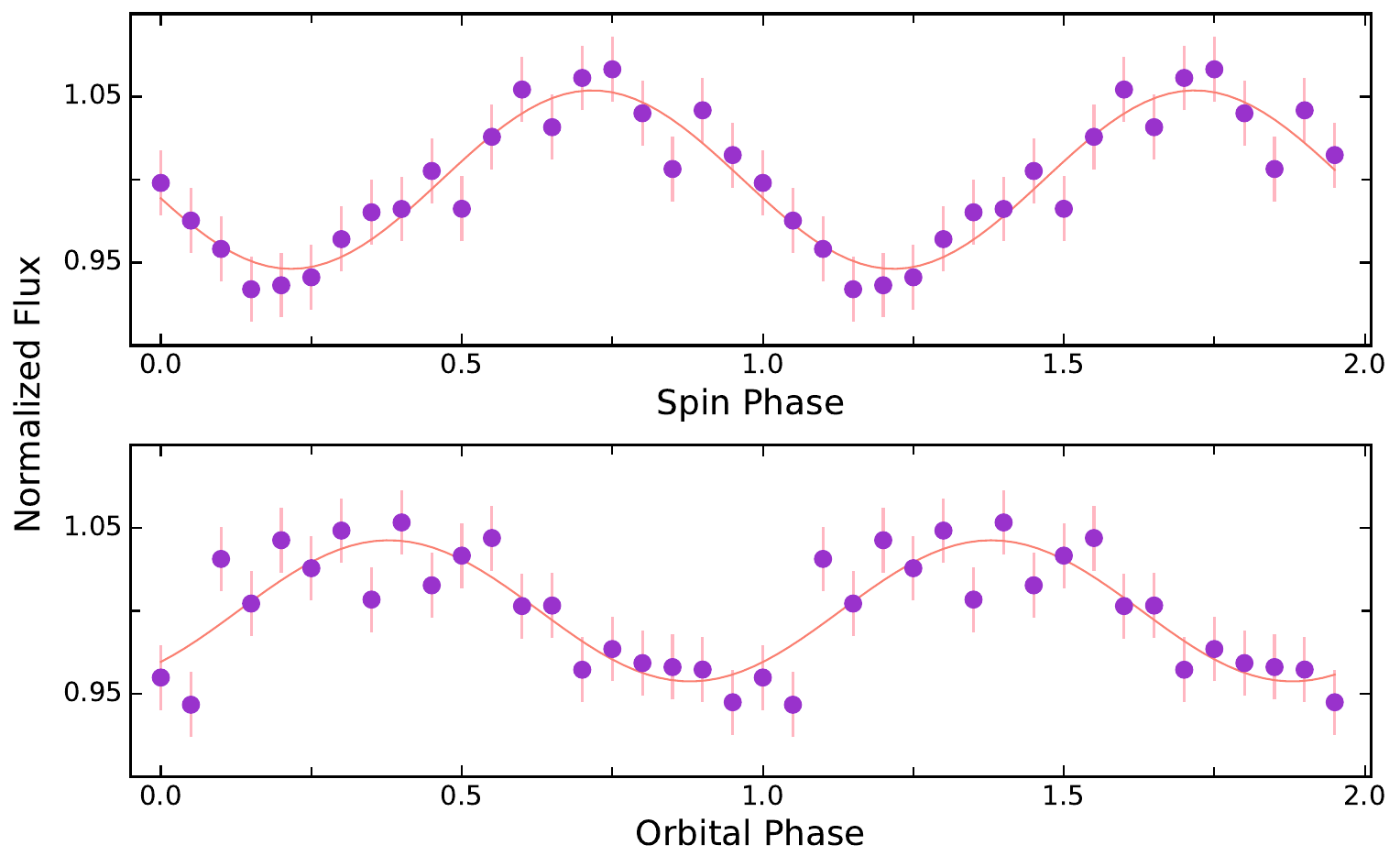}
\caption{Spin (top) and orbital (bottom) phase-folded \textit {TESS} light curves of J1616.}
\label{fig:tessorbspinflc_J1616}
\end{figure*}

\section{Analysis and results}\label{sec:analysis}


\subsection{2MASS J09213414$-$5939068} \label{sec:tesslcpsflc_J0921}
The \textit {TESS} light curves of J0921 for sectors 9 and 10 are shown in Figure \ref{fig:tesslc_J0921}. To search for the periodicity, a period analysis was conducted using the Lomb-Scargle (LS) periodogram algorithm \citep{Lomb76, Scargle82}. The LS power spectrum obtained from the combined \textit {TESS} observations of both sectors is shown in Figure \ref{fig:tessps_J0921}, where the positions of all identified frequencies are marked. The power spectrum displays a prominent peak that corresponds to a period of $\sim$ 908 s, which can be assigned as the spin period of the WD. The other significant peak at about a period of 1.5 hr corresponds to the second harmonic of the orbital period, precisely half of the orbital period of $\sim$ 3.04 hr, as determined by \cite{Pretorius08}. When I consider $\sim$ 908 s and $\sim$ 3.04 hr as the spin and orbital periods, respectively, the derived period at $\sim$ 990 s can be 
interpreted as the beat period of the system. The other identified periods correspond to the frequencies $\omega-2\Omega$ and $2$($\omega-\Omega$). These observed frequencies exhibit lower power values than the $\omega$ frequency. The significance of these detected peaks is determined by calculating the false-alarm probability \citep[see][]{Horne86}. The dashed horizontal black line represents the 90\% confidence level. Periods corresponding to these dominant peaks are also derived from sectors 9 and 10 separately. They are well consistent with the periods derived from the combined \textit {TESS} data and are given in Table \ref{tab:ps}. 
Based on the exceptionally long baseline provided by \textit{TESS}, the temporal evolution of the power spectra was also examined in successive one-day intervals. The data were segmented from each sector into successive intervals of one day. As a result, sectors 9 and 10 have a total of 24 and 22 data segments, respectively. Furthermore, the LS periodogram analysis was performed on each one-day dataset. Figures \ref{fig:1daypssec9_J0921} and \ref{fig:1daypssec10_J0921} exhibit one-day incremented trailed power spectra for sectors 9 and 10, respectively. The frequencies $2$$\Omega$, $\omega - \Omega$, and $\omega$ are irregular throughout the time-resolved power spectra and are sometimes significant and sometimes below the confidence level. In contrast to the combined power spectrum, no significant $\omega-2\Omega$ and $2$($\omega-\Omega$) signals are detected during one-day observations of either sector. 

The periodic variability of J0921 was also inspected by folding the light curves over the spin and orbital periods and using the first point of the \textit {TESS} observation as the reference epoch, BJD = 2458544.114307. Spin and orbital phase-folded light curves were obtained with a binning of 20 points in a phase along with the best-fit sine function and with the sum of two best-fit functions, a sine and a cosine, respectively. They are shown in Figure \ref{fig:tessorbspinflc_J0921}. 
Clear periodic sinusoidal modulations are observed in the spin-phase-folded light curve, whereas the orbital-phase-folded light curve exhibits double-peaked modulations. This is also evident from the power spectrum, which reveals a strong significant peak at the second harmonic of the orbital period. Furthermore, to show the evolution of the short-term variations, the day-wise periodic variation during the rotation of the WD was explored. Following the same approach as mentioned above, each day’s light curve was folded with a binning of 20 points in a phase. Figures \ref{fig:1dayspinflcsec9_J0921} and \ref{fig:1dayspinflcsec10_J0921} represent the colour-composite plots for the one-day spin-phased pulse profile for sectors 9 and 10, respectively. Strong single broad-peak spin modulations are observed throughout the one-day \textit {TESS} observations in both sectors.

\begin{figure*}
\centering
\subfigure[]{\includegraphics[width=0.82\textwidth]{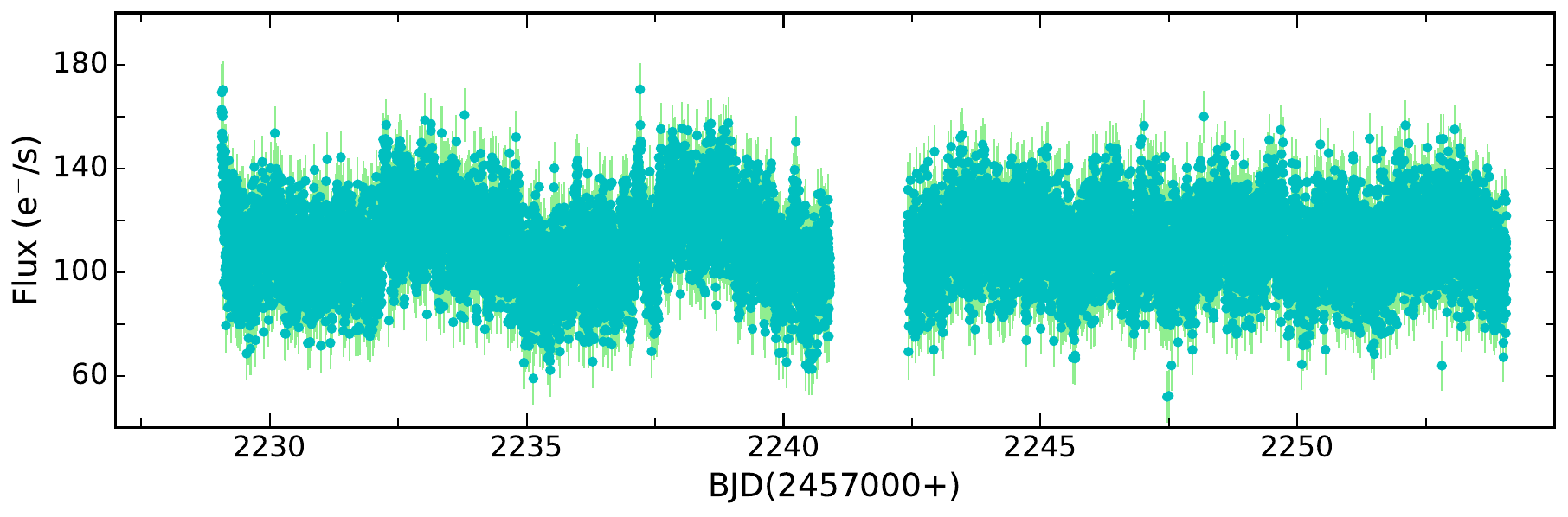}\label{fig:tesslc_v667pup}}
\subfigure[]{\includegraphics[width=0.82\textwidth]{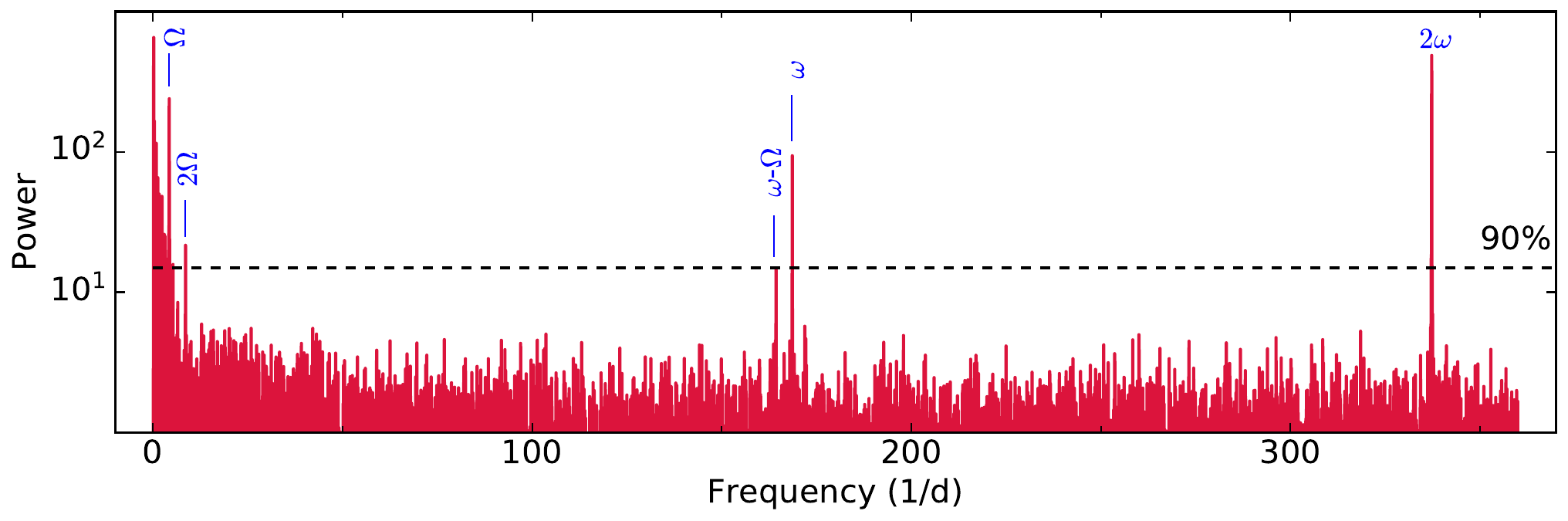}\label{fig:tessps_v667pup}}
\caption{(a) \textit{TESS} light curve  
 to reveal the brightness variations in V667 Pup, and (b) LS power spectrum obatined from the \textit {TESS} observations of V667 Pup. The significant frequencies observed in the power spectrum lie above the 90\% confidence level (represented by the dashed horizontal black line), and are marked.} 
\end{figure*}

\begin{figure*}
\centering
\hspace{1.5cm}
\includegraphics[width=0.82\textwidth]{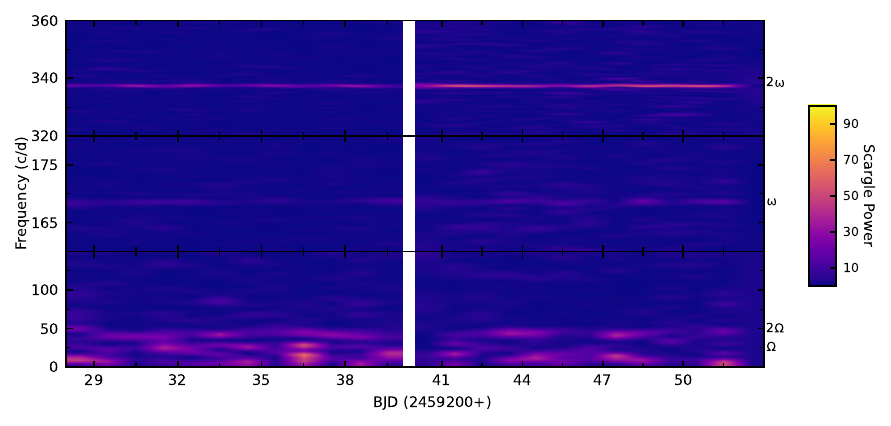}
\caption{One-day time-resolved \textit {TESS} power spectrum of V667 Pup near the $\Omega$, $\omega$, and $2\omega$ frequency regions.} 
\label{fig:1dayps_v667pup}
\end{figure*}

\begin{figure*}
\centering
\includegraphics[width=0.8\textwidth]{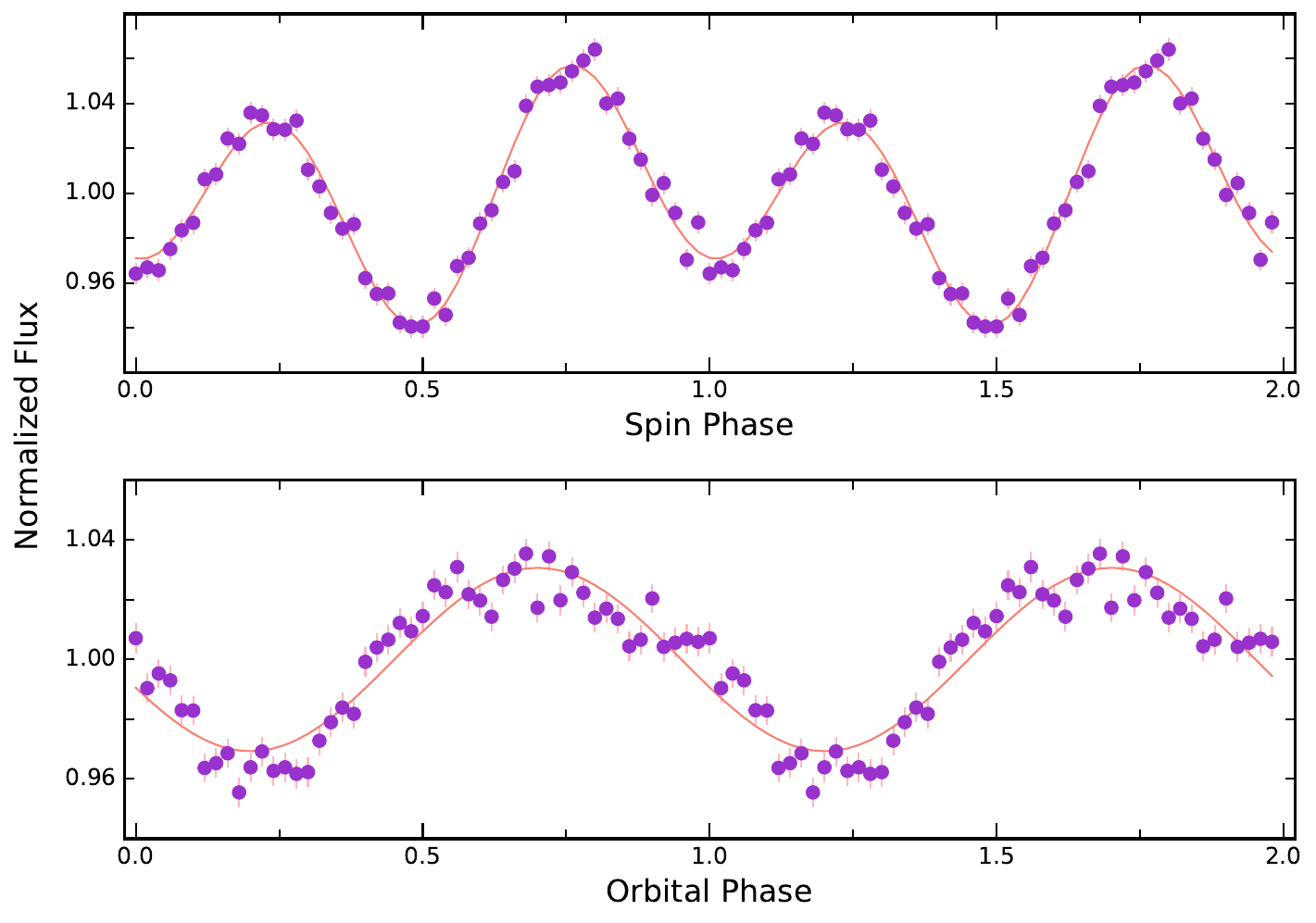}
\caption{Spin (top) and orbital (bottom) phase-folded \textit {TESS} light curves of V667 Pup.} 
\label{fig:tessorbspinflc_v667pup}
\end{figure*}

\begin{figure*}
\centering
\includegraphics[width=0.62\textwidth]{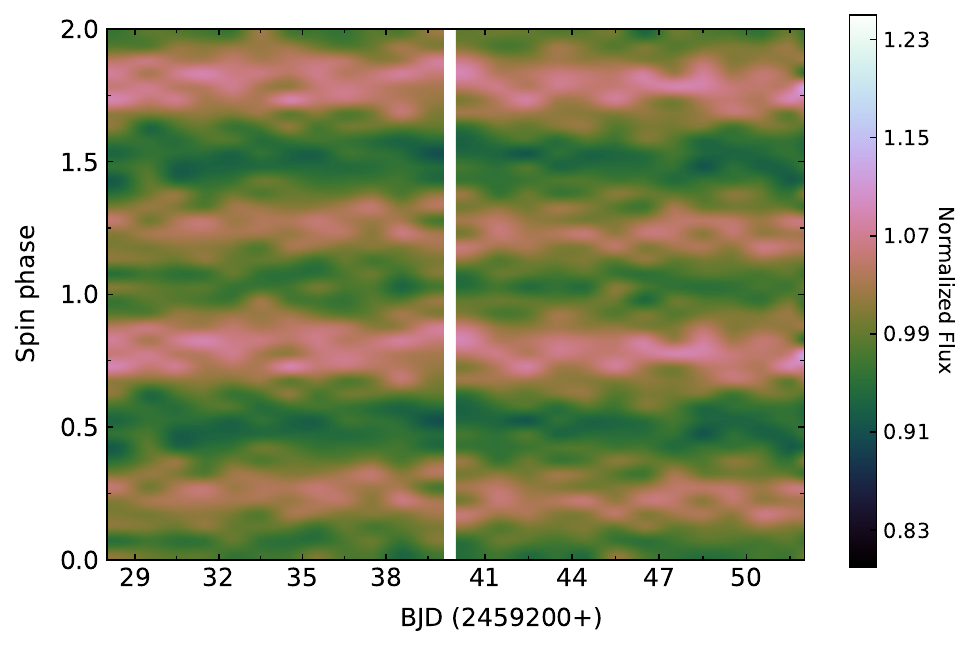}
\caption{Evolution of the spin-pulse profiles during one-day \textit {TESS} observations of V667 Pup.} 
\label{fig:1dayspinflc_v667pup}
\end{figure*}


\subsection{IGR J16167$-$4957} 
\label{sec:tesslcpsflc_J1616}
Figure \ref{fig:tesslc_J1616} displays the complete \textit {TESS} light curve of J1616, showing short-term fluctuations superimposed on long-term variability upon careful examination. In order to identify the periodicities, the LS periodogram analysis was performed on the entire \textit {TESS} dataset, which is shown in Figure \ref{fig:tessps_J1616}, where the positions of the two identified frequencies are marked. Both periods are observed above the 90\% confidence level. The detected significant frequency corresponds to the orbital period of 4.99$\pm$0.01 hr and confirms previous results of \cite{Pretorius09}. Another significant peak corresponds to the period of 582.45$\pm$0.04 s and can be considered as the spin period of the WD in J1616. The dominance of spin  modulation over orbital modulation is clearly visible in its power spectrum. Moreover, a cluster of frequencies in the lower-frequency region of 0.1 and 0.3 c/d  is also present in the power spectrum and is apparently unrelated to the frequencies identified in the J1616 power spectrum. Similar to J0921, a one-day time-resolved power spectrum analysis was also conducted for J1616. Unfortunately, the spin and orbital frequencies were both detected below the confidence level throughout each day-segment power spectrum. 

To study the phased-light curve variations, the \textit {TESS} light curve was folded using the zero time of the first \textit {TESS} observation, BJD=2458629.446496, and the derived orbital and spin periods. The light curves were folded with a binning of 20 points in a phase, and they are shown in Figure \ref{fig:tessorbspinflc_J1616}. The orbital and spin-phase-folded light curves exhibit sinusoidal modulations, and broad peaks occur in phases $\sim$ 0.35 and $\sim$ 0.75, respectively.


\subsection{V667 Pup} \label{sec:tesslcpsflc_V667pup}
Figure \ref{fig:tesslc_v667pup} shows the \textit {TESS} light curve of V667 Pup, where the variable nature of the source is evident. To probe the periodic nature of V667 Pup, the LS periodogram analysis was performed on the combined \textit {TESS} timing data. It is shown in Figure \ref{fig:tessps_v667pup}. Two fundamental periods at 5.58$\pm$0.01 hr and 512.40$\pm$0.03 s are observed in the LS power spectrum. They correspond to the orbital and spin periods of the system, respectively. The frequency corresponding to the period of 525.77$\pm$0.03 s is also present in the power spectrum and can be attributed to the beat period of the system, which was not detected in the earlier observations. All detected peaks lie above the 90\% confidence level. All the derived periods corresponding to five significant peaks at frequencies $\Omega$, $2\Omega$, $\omega$-$\Omega$, $\omega$, and $2\omega$ are given in Table \ref{tab:ps}. Similar to J0921, the evolution of the power spectra was also inspected in consecutive one-day time segments. 
The trailed LS power spectra with one-day increments are shown in Figure \ref{fig:1dayps_v667pup}. Each day-segment exhibits strong power only in the detected significant $2\omega$ frequency. However, the $\omega$ and $\Omega$ frequencies are sporadic throughout the time-resolved power spectra. No significant $2\Omega$ and $\omega-\Omega$ frequencies are detected in one-day power spectra, however. 

 The \textit {TESS} light curves were folded using the same reference epoch as given by \cite{Thorstensen13} in units of BJD as 2453812.7598 and more precise spin and orbital periods derived from the \textit {TESS} observations (see Table \ref{tab:ps}). The spin (along with the sum of two best-fit functions, a sine and a cosine) and orbital (along with a best-fit sine function) phase-folded light curves with a phase bin of 0.02 are shown in Figure \ref{fig:tessorbspinflc_v667pup}. The \textit {TESS} light curves reveal two prominent peaks during a rotation of the WD, where the amplitude of the first peak is slightly smaller than that of the second peak. However, a single broad hump is seen during half of the orbital cycle of V667 Pup. A similar behaviour of a double-peaked spin-pulse profile is also observed in the one-day \textit {TESS} observations, as depicted in Figure \ref{fig:1dayspinflc_v667pup}.


\section{Discussion}\label{sec:diss}
Detailed time-resolved analyses were carried out for three CVs, J0921, J1616, and V667 Pup, using high-cadence optical photometric data from the \textit {TESS}. I speculate that J0921 belongs to the IP class of magnetic CVs that accrete with the disc-overflow mechanism, whereas the detection of a spin signal in J1616 and a beat signal in V667 Pup supports the classification of these systems as IPs, where accretion may primarily be governed by disc-fed and disc-overflow mechanisms, respectively.

In the optical data from the \textit {TESS}, five significant periods were observed in J0921, including a fundamental period of 908.12$\pm$0.05 s, which is probably the spin period of the WD. Considering $\sim$ 908 s and $\sim$ 3.04 hr as spin and orbital periods, respectively, I estimated the beat period to be $\sim$ 990 s, which is also present in the power spectrum of J0921. The presence of these frequencies indicates that J0921 may belong to the IP class of magnetic CVs. The dominant peak at the $2\Omega$ frequency may suggest a strong contribution from the secondary star due to ellipsoidal modulation. The combinations of spin and orbital frequencies are typically considered the primary frequency components in the search for a possible accretion mechanism in IPs. In this context, the detection of $\omega$ $-$ $ \Omega$ in J0921 seems intrinsic and provides indications that there would be material flow with an accretion stream. This modulation is  not entirely caused by the amplitude modulation of the spin frequency at the orbital period: If it were the orbital modulation of the $\omega$ frequency, then $\omega$ $+$ $\Omega$ should be present in the power spectrum with the same power as $\omega$ $-$ $\Omega$, which is not observed in our data. 
 Moreover, the origin of the $\omega-2\Omega$ freqeuncy cannot be due to the modulation of $\omega$ at $2\Omega$; otherwise, $\omega+2\Omega$ should also be present in the power spectrum. The simultaneous presence of $\omega-2\Omega$ along with the $\omega-\Omega$ frequency confirms that the origin of $\omega-2\Omega$ is due to the orbital modulation of the $\omega-\Omega$ component ($\omega-\Omega$ $\pm$ $\Omega$=$\omega-2\Omega$ and $\omega$), as was pointed out by \cite{Warner86}.  

The detection of $\omega-\Omega$ along with strong $\omega$ indicates that J0921 may accrete via a combination of disc and stream. Additionally, one-day time-resolved power spectra revealed fluctuations in the significant powers of $\omega$ and $\omega-\Omega$. In sector 10, particularly on day 73, the power of $\omega-\Omega$ is notably higher than the power of $\omega$, indicating the dominance of stream-fed accretion. However, for the remaining observations, the power associated with $\omega-\Omega$ is either undetected or is relatively lower than $\omega$, which suggests that J0921 likely accretes predominantly through the disc or accretes via the disc-overflow mechanism with disc-fed dominance. The observed change may be attributed to variations in both the mass accretion rate and the activity of the secondary star. In earlier studies, the alterations in the accretion mechanism were also investigated by examining the power spectra of IPs, focusing on the influence of the spin and beat frequency \citep[][]{Buckley89, Norton97, Littlefield20, Rawat21}. Thus, a variable disc-overflow accretion may be one of the possible accretion mechanisms during the \textit{TESS} observations of J0921. Alternatively, \cite{Murray99} suggested that a spiral disc structure might give rise to modulations at orbital sidebands of the spin frequency in IPs, including a beat and its second harmonic. Moreover, \cite{Warner86} pointed out that the reprocessing of X-rays by structures stationary in the binary rest frame would yield an optical beat signal in IPs. These could also provide alternative explanations for the detection of the beat signal in J0921.

For J1616, a significant period of 4.99$\pm$0.01 hr is detected, which corresponds to the orbital period of the system. Due to the better time-cadence and longer-baseline of \textit {TESS}, a clearly significant period of 582.45$\pm$0.04 s is observed in its power spectrum. No clear detection of this period was observed previously by \cite{Pretorius09} and \cite{Butters11}. However, within the uncertainties, this 582 s signal may match the intermittent 585 s signal observed by \cite{Pretorius09}. The detected period of 582.45$\pm$0.04 s appears to be the probable spin period of the WD in J1616. Thus, the existence of dominant modulation at this freqeuncy  supports its classification as an IP. The IPs are generally united with three groups: slow rotators with $P_\omega$/$P_\Omega$ $\sim$ 0.5, intermediate rotators with $P_\omega$/$P_\Omega$ $\sim$ 0.1, and fast rotators with $P_\omega$/$P_\Omega$ $\sim$ 0.01. If the proposed spin period is indeed the actual period, then  $P_\omega$/$P_\Omega$ $\sim$ 0.03 for J1616. Thus, with $P_\omega$/$P_\Omega$ $\sim$ 0.03 or $P_\omega$ $\sim$ 582 s, J1616 appears to belong to the fast-rotator category. With these values of the orbital and spin periods, the inferred value of $P_{\omega-\Omega}$ period is $\sim$ 601.96 s, which corresponds to the frequency of $\sim$ 143.5 c/d. No period corresponding to this frequency is detected in its power spectrum. The dominant $\omega$ and the absence of $\omega-\Omega$ or any other sideband frequency signifies that it might be a disc-fed accretor, however. Nevertheless, further follow-up X-ray observations are needed to better characterise the candidate spin signal and the accretion geometry of the system.
 
Five significant frequencies, $\Omega$, $2\Omega$, $\omega-\Omega$, $\omega$, and $2\omega$, are detected in the optical \textit {TESS} power spectrum of V667 Pup. All the observed frequencies reinforce the IP nature of V667 Pup. Based on the longer baseline available with the \textit {TESS} data, a refined spin period of 512.40$\pm$ 0.03 s is detected. This derived spin period is found to be clearly consistent with the periods derived from previous measurements \citep[see][]{Patterson06, Butters07}. The \textit {TESS} data derived the orbital period of V667 Pup as 5.58$\pm$0.01 hr, which is found to be somewhat inconsistent with the spectroscopic period of 5.611$\pm$0.005 determined by \cite{Thorstensen13}. With P$_\omega$/P$_\Omega$ $\sim$ 0.03 and P$_\omega$ $\sim$ 512 s, V667 Pup also seems to belong to the category of fast rotators, similar to the other IPs, RX J2133.7+5107, V455 And, V418 Gem, and NY Lup. Along with the spin and orbital periods, the other observed frequencies can also serve as an important diagnostic to understand the mode of accretion in this system. Among them, a significant $\omega-\Omega$ frequency with a period of 525.77$\pm$0.03 s is detected in the study of the V667 Pup for the first time. Similar to J0921, the presence of $\omega-\Omega$ seems intrinsic because $\omega+\Omega$ is not detected in its power spectrum, which safely affirms that the detection of the beat modulation is entirely intrinsic and originated from accretion through the stream. Although the strong $\omega$ and $2\omega$ frequencies suggest the dominance of disc-fed accretion, however, the detection of an additional beat frequency implies that some fraction of material is also accreted via a stream. Therefore, V667 Pup is also likely to be a disc-overflow accreting intermediate polar.

The observed optical spin modulations in V667 Pup are non-sinusoidal compared to most IPs and exhibit a double-peaked pulse profile, as expected from the observed second harmonic of the spin pulse in the power spectrum, which can be interpreted in terms of two emitting poles. The relative contribution of the two poles to the observed emission can be explained based on the ratio of the fundamental-to-second harmonic amplitude. The amplitude of the first hump is found to be smaller than that of the second hump, and both peaks are separated by about 0.5 in phase, indicating that the two poles may accrete at a different rate and that they are separated by 180$^\circ$. Based on the strength of the magnetic field of the WD, \cite{Norton99} suggested that the two-pole accretion model can describe the double-peaked profile structure in IPs. Their model predicts that weak magnetic fields are typical for short-period spinning IPs, which results in double-peaked pulse profiles. This might be one of the reasons that can explain the double-peaked optical pulse profile in the short-period rotating IP V667 Pup. Another possibility could be that the geometry allows the two opposite poles to come into view for the observer \citep[e.g., V405 Aur][]{Evans04}. This means that the geometry of the system might be such that at certain  orientations, both poles (which are located on opposite sides of the system) are aligned in a way that they can be seen simultaneously by the observer. This alignment allows us to observe the distinct double-peaked pattern in the optical pulse. Other than the spin modulations, obscuration of the WD by the material rotating in the binary frame or an eclipse of the hotspot by an optically thick disc may cause the broad single-peaked orbital modulation in V667 pup.


\section{Conclusions}\label{sec:conc}
The characteristics of J0921, J1616, and V667 Pup are summarised below.
\begin{enumerate}
\item J0921 is probably an intermediate polar with a detection of two periods at 908.12$\pm$0.05 s and 990.10$\pm$0.06 s, which are interpreted as the spin and beat periods, respectively. The detection of both $\omega-\Omega$ and $\omega$ suggests that J0921 is likely undergoing through the disc-overflow mechanism.
\item The detection of an unambiguous dominant period at 582.45$\pm$0.04 s in J1616 can be interpreted as the spin period of the WD, which supports its classification as an IP, where accretion may primarily be governed by a disc.
\item Refined spin and orbital periods of 512.40$\pm$0.03 s and 5.58$\pm$0.01 hr, respectively, were derived using the \textit{TESS} observations of V667 Pup. They further confirm its IP nature. However, an unambiguous  beat period of 525.77$\pm$0.03 s is detected for the first time in the study of this system.
\item The strong $\omega$ and $2\omega$ frequencies suggest that V667 Pup might be accreting predominantly via the disc. The detection of a previously unknown $\omega-\Omega $ frequency indicates that part of the accreting material also directly flows towards the WD along the magnetic field lines, however. 
\item The double-peaked structure with varying amplitudes observed in the optical spin pulse profile of V667 Pup suggests a possible two-pole accretion geometry, where each pole accretes at a different rate and is separated by 180$^\circ$.
\end{enumerate}



\section*{Acknowledgments}
I thank the anonymous referee for providing useful comments and suggestions that improved the manuscript considerably. AJ acknowledges support from the Centro de Astrofisica y Tecnologias Afines (CATA) fellowship via grant Agencia Nacional de Investigacion y Desarrollo (ANID), BASAL FB210003. This paper includes data collected with the \textit {TESS} mission, obtained from the MAST data archive at the Space Telescope Science Institute (STScI). Funding for the \textit {TESS} mission is provided by the NASA Explorer Program. 


\bibliographystyle{aa}
\bibliography{ref}
 
\end{document}